\documentstyle[amssymb,preprint,aps]{revtex}

\begin{document}
\title{A generalized molecular theory for nematic liquid crystals formed by
non-cylindrically symmetric molecules}
\author{O. Kayacan\thanks{
e-mail: ozhan.kayacan@bayar.edu.tr}}
\address{Physics Department, Art and Science Faculty, Celal Bayar University, {%
Muradiye/Manisa-TURKEY}}
\maketitle

\begin{abstract}
\noindent Many molecular theories of nematic liquid crystals consider the
constituent molecules as cylindrically symmetric. In many cases, this
approximation may be useful. However the molecules of real nematics have
lower symmetry. Therefore a theory was developed (Mol. Phys. 30 (1975) 1345)
for an ensemble of such particles based upon a general expansion of the
pairwise intermolecular potential together with the molecular field
approximation. In this study, we would like to handle this molecular field
theory by using Tsallis thermostatistics which has been commonly used for a
decade to study the physical systems. With this aim, we would like to
investigate the dependence of the order parameters on temperature and would
like to report the variation of the critical values of the order parameters
at the transition temperature with the entropic index.

{PACS Number(s):}~ 05.20.-y, 05.70.-a, 61.30.Cz, 61.30. Gd.\newline

\end{abstract}

\newpage

\section{Introduction}

It is well known that an essential characteristic of compounds forming
liquid crystals is the rod-like shape of their constituent molecules, with
an high length to breadth ratio. Therefore the molecules are supposed to be
cylindrically symmetrical. For example, the ordering matrix which is often
used to describe the partial alignment in a mesophase, contains only one
independent element and this can be determined by some techniques [1]. The
fact that the molecular cylindrical symmetry is assumed is appealing to a
statistical mechanician, because the pairwise anisotropic intermolecular
potential required in any calculation is simple for such particles [2].
However the molecules, in fact, are lath-like and thus do not possess the
high symmetry.

The ordering matrix has two principal components and therefore these
components are required to describe the orientational order of a uniaxial
mesophase composed of lath-like molecules. In this sense, the deviation of
this ordering matrix from cylindrical symmetry was found to be significant
[3]. The importance of deviations from cylindrical symmetry may be inferred
from unambiguous determinations of the ordering matrix for rod-like
molecules, such as $4,4^{\prime }-dichlorobiphenyl$ [4]. Moreover it is
found that these matrices are comparable to those estimated for a pure
mesophase [3].

There are some studies in which the consequences of deviations from
molecular cylindrical symmetry are investigated. It is shown that a system
consisting of particles with a lower symmetry than $D_{\infty h}$ is capable
of existing either as a uniaxial or a biaxial liquid crystal [5]. The
possible existence of a biaxial phase is studied in detail for a system of
hard rectangular plates using a lattice model [6], the Landau approach [7]
and the molecular field approximation [8]. The deviations of the ordering
matrix from cylindrical symmetry is clearly determined by the molecular
symmetry and the element of the ordering matrix for the long axis will also
be influenced by the form of the pseudo-intermolecular potential. The
calculations of the ordering matrix for an ensemble of hard rectangular
particles is performed in [9]. It must be emphasized that although these
calculations are of some interest, they may not be particularly realistic
because some experiments indicate that dispersion forces may make a dominant
contribution to the anisotropic intermolecular potential [9]. Considering
the cases above, Luckhurst et al. developed a theory [10] for
non-cylindrically symmetric molecules interacting via a completely general
intermolecular potential within molecular field approximation.

For a decade, nonextensive statistics has an increasing interest and
recently Tsallis thermostatistics (TT) has been applied to the
nematic-isotropic transition [11-13] as a nonextensive statistics. In [11],
the Maier-Saupe mean field theory has been generalized within TT and applied
to a nematic liquid crystal, para-azoxyanisole. In the other study, [12],
the the effects of the nonextensivity on the dimerization process has been
studied, and finally the mean field theory of anisotropic potentail of rank $%
L=4$ has been generalized within TT and the effect of the nonextensivity on
the order parameters has been illustrated in [13]. Up to now, the mean field
theories for uniaxial nematogens formed by cylindrically symmetric molecules
have been studied by using TT. In this manner, we aim, in this study, to
enlarge the applications of TT to the liquid crystal systems and to handle
Luckhurst et al.'s theory which considers the molecules to be
non-cylindrically symmetric. In doing so, we first give some essential
properties of Luckhurst et al.'s theory. Then we mention on TT and its
axioms. Finally, we apply TT to the Luckhurst et al.'s theory and some
possible concluding remarks are made. We must emphasize that we would like
to give only the possible contributions of the nonextensivity to the theory.
So we must keep in mind that since one relies on the generalized theory or
not, more extensional studies related with it must be performed in the
nematic-isotropic transition. However, we believe that this study is
sufficient to give motivation for further applications of TT to the liquid
crystals.

\subsection{The Molecular Theory for Uniaxial Nematics Formed by
Non-cylindrically Symmetric Molecules}

The intermolecular potential for particles of general shape is given by [10] 
\begin{equation}
U_{12}=\sum
u_{L_{1}L_{2}m_{1}m_{2}n_{1}n_{2}}(r_{12})\;D_{m_{1},n_{1}}^{(L_{1})}(\Omega
_{1-R})\;D_{m_{2},n_{2}}^{(L_{2})}(\Omega _{2-R})
\end{equation}
in a product basis of Wigner rotation matrix [14], where $r_{12}$ is the
distance between molecules $1$ and $2$. The orientation of molecule $\Omega
_{1-R}$ in a coordinate system containing the intermolecular vector as the $%
z $ axis is denoted by $\Omega _{i-R}$. This potential energy is invariant
under rotation of the coordinate system about $z$ axis. Therefore the
summation in Eq.(1) can be restricted \ as follows [10]: 
\begin{equation}
U_{12}=\sum u_{L_{1}L_{2}mn_{1}n_{2}}(r_{12})\;D_{m,n_{1}}^{(L_{1})}(\Omega
_{1-R})\;D_{-m,n_{2}}^{(L_{2})}(\Omega _{2-R}).
\end{equation}
In what follows, the redundant subscripts on the coefficient $u(r_{12})$
will be suppressed. Because it is convenient to define the molecular
orientation in terms of a common coordinate system, the potential energy $%
U_{12}$ could be transformed to a laboratory frame. The choice of this
coordinate system is determined according to the symmetry of the liquid
crystal phase, so for a uniaxial mesophase, the laboratory $z$ axis can be
taken to be parallel to the symmetry axis of the mesophase. The
transformation of $U_{12}$ is carried out by performing the rotation from
the intermolecular vector to the molecule in two steps using the
relationship 
\begin{equation}
D_{m,n}^{(L)}(\Omega _{1-R})=\sum_{j}D_{j,m}^{(L)\ast }(\Omega
_{R-L})\;D_{j,n}^{(L)}(\Omega _{1-R}),
\end{equation}
where the subscript $R-L$ is the rotation from the laboratory to the
intermolecular frame, $1-L$ denotes that from the laboratory to the molecule
coordinate system. Then the intermolecular potential can be written as 
\begin{equation}
U_{12}=\sum
u_{L_{1}L_{2}mn_{1}n_{2}}(r_{12})\;D_{j_{1},n_{1}}^{(L_{1})}(\Omega
_{1-L})\;D_{q_{2},n_{2}}^{(L_{2})}(\Omega
_{2-L})\,\,D_{j_{1},m}^{(L_{1})^{\ast }}(\Omega
_{R-L})\;D_{j_{2},-m}^{(L_{2})^{\ast }}(\Omega _{R-L}).
\end{equation}
If the distribution function for the intermolecular vector is independent of
orientation, then one could use the orthogonality of the rotation matrices
to evaluate the ensemble average: 
\begin{equation}
\overline{D_{j_{1},m}^{(L_{1})^{\ast }}(\Omega
_{R-L})\;D_{j_{2},-m}^{(L_{2})^{\ast }}(\Omega _{R-L})}=\frac{\left(
-1\right) ^{j_{2}+m}\delta _{j_{1},-j_{2}}\delta _{L_{1},L_{2}}}{2L_{1}+1}.
\end{equation}
The partially averaged potential may then be written as 
\begin{equation}
U_{12}=\sum \frac{\left( -1\right) ^{j+m}}{2L+1}%
u_{Lmn_{1}n_{2}}(r_{12})D_{j,n_{1}}^{(L)}(\Omega
_{1-L})D_{-j,n_{2}}^{(L)}(\Omega _{2-L}).
\end{equation}
Then now it is the time to average over the orientations adopted by particle 
$2$. However one needs $\overline{D_{-j,n_{2}}^{(L)}}$. This average is
taken to be independent of the orientation of molecule $1$ within the
molecular field approximation and it is to be identified with the normal
ensemble average. Since we only consider a uniaxial mesophase, $\overline{%
D_{-j,n_{2}}^{(L)}}$ vanishes when $j\neq 0$ and $L$ is even. In this
manner, the potential is reduced to the following form: 
\begin{equation}
U_{12}=\sum_{L,even}\frac{\left( -1\right) ^{m}}{2L+1}%
u_{Lmn_{1}n_{2}}(r_{12})D_{0,n_{1}}^{(L)}(\Omega _{1-L})\overline{%
D_{0,n_{2}}^{(L)}(\Omega _{2-L})},
\end{equation}
where the summation is over $L\geqslant 0$ ($L$ even). The scalar
contribution could be ignored because we would only like to study the
orientational properties of the mesophase. The average of the orientational
pseudo-potential for molecule $1$ is given by 
\begin{equation}
U_{1}=\sum_{L,even}\frac{\left( -1\right) ^{m}}{2L+1}\overline{u}%
_{Lmn_{1}n_{2}}D_{0,n_{1}}^{(L)}(\Omega _{1-L})\overline{D_{0,n_{2}}^{(L)}}.
\end{equation}
The expansion coefficients in Eq.(8) are defined by 
\begin{equation}
\overline{u}_{Lmn_{1}n_{2}}=\frac{\int u_{Lmn_{1}n_{2}}(r_{12})\;\exp \left(
-U_{o}({\bf r}_{1}...{\bf r}_{N})/kT\right) \;{\bf dr}_{1}...{\bf dr}_{N}}{%
\int \exp \left( -U_{o}({\bf r}_{1}...{\bf r}_{N})/kT\right) \;{\bf dr}%
_{1}...{\bf dr}_{N}},
\end{equation}
where $U_{o}({\bf r}_{1}...{\bf r}_{N})$\ is the scalar potential of the
ensemble [10,15].

The pseudo-potential is conveniently written as 
\begin{equation}
U\left( \beta \gamma \right) =\sum_{L,even}c_{Ljp}\,\overline{D_{0,p}^{(L)}}%
\,D_{0,j}^{(L)}\left( \beta \gamma \right) ,
\end{equation}
where 
\begin{equation}
c_{Ljp}=\sum_{m}\left( -1\right) ^{m}\frac{\overline{u}_{Lmjp}}{2L+1},
\end{equation}
with $\beta $ and $\gamma $ is the Euler angles and define the orientation
of the director in the molecular coordinate system.

Luckhurst et al. made some comments in [10] on this form of the
pseudo-potential and the symmetry properties of the coefficients $c_{Ljp}$.
Because Luckhurst et al. investigated the influence of deviations from
molecular cylindrical symmetry on various orientational order parameters for
a uniaxial nematic mesophase, they tried to minimize the number of the
variables in the calculation without any loss of the essential physics. As a
first approximation, they considered only those terms with $L$ equal to $2$
in the expansion of the pseudo-potential. The neglect of terms higher than
quadratic could be a good approximation, since similar assumptions for
cylindrically symmetric molecules had let to results in reasonable accord
with experiment [10,16]. As mentioned in [10], the number of expansion
coefficients are reduced by appealing to some specific model for the
interactions or by imposing symmetry restrictions on the molecules. If one
follows Straley [9] and takes the molecules to be hard rectangular
parallelopipeds, then the coefficients $c_{2jp}$\ are independent of the
sign of either $p$ or $j$ and zero if either of these subscripts is odd,
that is 
\begin{eqnarray}
c_{200} &=&\left[ -2B\left( W^{2}+L^{2}\right) -2W\left( L^{2}+B^{2}\right)
+L\left( W^{2}+B^{2}\right) +8WBL\right] /3, \\
c_{220} &=&\left( L^{2}-BW\right) \left( B-W\right) /\sqrt{6}, \\
c_{222} &=&-L\left( W-B\right) ^{2}/2,
\end{eqnarray}
where $L$ is the length, $B$ the breadth and $W$ the width of the molecule.
It must be noted that this parametrization is only approximate [9] and also
anisotropic repulsive forces are probably not dominant in determining the
behaviour of real nematics. Therefore assuming the molecules to interact via
dispersion forces, the expressions for $c_{2jp}$ could be derived.

The similar restriction can be imposed on the expansion coefficients $%
c_{2jp} $, without appealing to specific forms of the intermolecular
interactions, by using more formal arguments based on the molecular symmetry
and its influence on the pair potential. For example, if each molecule has a
centre of symmetry, then [10] 
\begin{eqnarray}
u_{L_{1}L_{2}mn_{1}n_{2}}(r_{12}) &=&\left( -1\right)
^{L_{1}-n_{1}}u_{L_{1}L_{2}m-n_{1}n_{2}}(r_{12})=\left( -1\right)
^{L_{2}-n_{2}}u_{L_{1}L_{2}mn_{1}-n_{2}}(r_{12})  \nonumber \\
&=&\left( -1\right)
^{L_{1}+L_{2}-n_{1}-n_{2}}u_{L_{1}L_{2}m-n_{1}-n_{2}}(r_{12})
\end{eqnarray}
and consequently 
\begin{equation}
c_{Ljp}=\left( -1\right) ^{j}c_{L-jp}=\left( -1\right) ^{p}c_{Lj-p}=\left(
-1\right) ^{j+p}c_{L-j-p}.
\end{equation}
In addition, if the molecules also possess a plane of symmetry orthogonal to
their $z$ axes, then 
\begin{eqnarray}
u_{L_{1}L_{2}mn_{1}n_{2}}(r_{12}) &=&\left( -1\right)
^{L_{1}}u_{L_{1}L_{2}m-n_{1}n_{2}}(r_{12})=\left( -1\right)
^{L_{2}}u_{L_{1}L_{2}mn_{1}-n_{2}}(r_{12})  \nonumber \\
&=&\left( -1\right) ^{L_{1}+L_{2}}u_{L_{1}L_{2}m-n_{1}-n_{2}}(r_{12})
\end{eqnarray}
and 
\begin{equation}
c_{Ljp}=\left( -1\right) ^{L}c_{L-jp}=\left( -1\right)
^{L}c_{Lj-p}=c_{L-j-p}.
\end{equation}
However Eqs.(16,18) can only be consistent when both $p$ and $j$ have even
values. Moreover the coefficients are independent of the sign of $p$ and $j$.

Then the pseudo-potential may be written as 
\begin{equation}
U\left( \beta \gamma \right) =\sum_{L,\left| p\right| ,\left| j\right|
\;(even)}c_{Ljp}\left[ D_{0,p}^{(L)}+D_{0,-p}^{(L)}\right] \left[
D_{0,j}^{(L)}(\beta \gamma )+D_{0,-j}^{(L)}(\beta \gamma )\right] /(1+\delta
_{0j})(1+\delta _{0p}).
\end{equation}
according to the restrictions above. If limiting the summation to those
terms with $L$ equal to $2$, one has 
\begin{eqnarray}
U\left( \beta \gamma \right) &=&\left[ c_{200}\overline{d_{0,0}^{(2)}}%
+2c_{220}\overline{d_{0,2}^{(2)}\cos (2\gamma )}\right] d_{0,0}^{(2)}(\beta )
\nonumber \\
&&+2\left[ c_{200}\overline{d_{0,0}^{(2)}}+2c_{222}\overline{d_{02}^{(2)}}%
\cos (2\gamma )\right] d_{0,2}^{(2)}(\beta )\cos (2\gamma ),
\end{eqnarray}
where $d_{m,n}^{(2)}(\beta )$ is a reduced Wigner rotation matrix [14]. The
second rank order parameters $\overline{d_{0,0}^{(2)}}$ and $\overline{%
d_{0,2}^{(2)}cos(2\gamma )}$ in this expression are related to the principal
elements of the ordering matrix by 
\begin{equation}
S_{zz}=\overline{d_{0,0}^{(2)}}
\end{equation}
and 
\begin{equation}
S_{xx}-S_{yy}=\sqrt{6}\overline{\;d_{0,2}^{(2)}\cos (2\gamma )}.
\end{equation}
The order parameter $\overline{d_{0,2}^{(2)}\cos (2\gamma )}$ indicates the
deviation of the ordering matrix from cylindrical symmetry. For simplicity,
it is convenient to write the pseudo-potential as 
\begin{equation}
U(\beta \gamma )=-kT\left[ a\,d_{0,0}^{(2)}(\beta )+b\,d_{0,2}^{(2)}(\beta
)\cos (2\gamma )\right] ,
\end{equation}
where 
\begin{equation}
a=-\left[ c_{200}\overline{\;d_{0,0}^{(2)}}+2\,c_{220}\overline{%
\;d_{0,2}^{(2)}\cos (2\gamma )}\right] /kT
\end{equation}
and 
\begin{equation}
b=-\left[ 2\,c_{220}\;\overline{d_{0,0}^{(2)}}+4\,c_{222}\overline{%
\;d_{0,2}^{(2)}\cos (2\gamma )}\right] /kT.
\end{equation}
The partition function for this single particle potential is then given by 
\begin{equation}
Z=\int_{0}^{2\pi }\int_{0}^{\pi }\exp \left[ a\;d_{0,0}^{(2)}(\beta
)+b\;d_{0,2}^{(2)}(\beta )\cos (2\gamma )\right] \sin \beta d\beta \,d\gamma
.
\end{equation}
The orientational molar potential energy and the molar entropy are given as
follows 
\begin{eqnarray}
U &=&-(RT/2)\left[ a\,\overline{d_{0,0}^{(2)}}+b\,\overline{%
d_{0,2}^{(2)}\cos (2\gamma )}\right] , \\
S &=&-R\left[ a\,\overline{d_{0,0}^{(2)}}+b\,\overline{d_{0,2}^{(2)}\cos
(2\gamma )}\right] +R\ln (Z).
\end{eqnarray}
respectively. Then the orientational molar free energy is straightforward: 
\begin{equation}
F_{m}=\frac{RT}{2}\left[ a\,\overline{d_{0,0}^{(2)}}+b\,\overline{%
d_{0,2}^{(2)}\cos (2\gamma )}\right] -RT\ln (Z).
\end{equation}
The order parameters are calculated from the equations below: 
\begin{equation}
\overline{d_{0,0}^{(2)}}=\frac{\int_{0}^{2\pi }\int_{0}^{\pi
}d_{0,0}^{(2)}(\beta )\;\exp \left[ ad_{0,0}^{(2)}(\beta
)+bd_{0,2}^{(2)}(\beta )\cos (2\gamma )\right] \sin \beta d\beta \,d\gamma }{%
Z}
\end{equation}
and 
\begin{equation}
\overline{d_{0,2}^{(2)}\cos (2\gamma )}=\frac{\int_{0}^{2\pi }\int_{0}^{\pi
}d_{0,2}^{(2)}(\beta )\cos (2\gamma )\;\exp \left[ ad_{0,0}^{(2)}(\beta
)+bd_{0,2}^{(2)}(\beta )\cos (2\gamma )\right] \sin \beta d\beta \,d\gamma }{%
Z}.
\end{equation}
It is worthwhile to note that $\overline{d_{0,0}^{(2)}}$ in Eq.(30) is
simply $\overline{P_{2}}$, ($P_{2}$ is the second Legendre polynomial). The
free energy $F_{m}$ will be a minimum provided the consistency equations for
these order parameters are satisfied. With the aid of Eqs.(24,25), one can
write the following equations: 
\begin{equation}
a=-\alpha \left[ \overline{d_{0,0}^{(2)}}+2\lambda \overline{%
\;d_{0,2}^{(2)}\cos (2\gamma )}\right] ,
\end{equation}
and 
\begin{equation}
b=2\lambda a,
\end{equation}
where 
\begin{equation}
\alpha =-c_{200}/kT.
\end{equation}
$\lambda $ in Eq.(34) is the ratio $c_{220}/c_{200}$ and it is a measure of
the deviation from cylindrical symmetry. If $\lambda $ is taken to be zero,
the theory reduced to the Maier-Saupe mean field theory.

Now we would like to give the description of Tsallis thermostatistics with
its axioms below.

\subsection{Generalization of the theory}

Tsallis thermostatistics (TT) has been extensively used to investigate the
concepts of statistical mechanics and thermodynamics. In this context, it
has been applied to various phenomena [17] and TT seems to be appropriate
for the study of nonextensive systems. As mentioned above, we have choosen
the nematic-isotropic transition as an application area of TT in earlier
studies [11-13] and the present study would be the extensional study to
investigate the nematic liquid crystals. In all of these earlier studies, we
consider that the uniaxial nematic liquid crystals are formed by
cylindrically symmetric molecules. However, we wonder in this study that if
the mesophase could be formed by non-cylindrically symmetric molecules, how
does the nonextensivity affect the nematic-isotropic transition? Therefore
our starting point is Luckhurst et al.'s theory [10] which is a molecular
field theory and assumes the molecules of the nematic liquid crystals to be
non-cylindrical and we examine the effects of the nonextensivity on the
nematic-isotropic transition by considering this molecular field theory.

The understanding of the basic principles and properties of TT has gain
fundamental importance. In this manner, Tsallis et al. have studied the role
of the constraint within TT. Tsallis proposed an entropy definition in 1988
[18]: 
\begin{equation}
S_{q}=k\frac{1-\sum_{i=1}^{W}p_{i}^{q}}{q-1}\;\;\;\left( q\in \Re \right) ,
\end{equation}
where $k$ is a constant, $p_{i}$ is the probabilty of the system in the $i$
microstate, $W$ is the total number of configurations and $q$ is called the
entropic index whose meaning will be given below. In the limit $q\rightarrow
1$, the entropy defined in Eq.(35) reduced to the Boltzmann-Gibbs (BG)
entropy. In other words, TT contains Boltzmann-Gibbs statistics as a special
case with $q=1$. As well known, BG statistics is a powerful one to study a
variety of the physical systems. However, it fails for the systems which $%
(i) $ have long range interactions, $(ii)$ have long range memory effects
and $(iii)$ evolve in multi-fractal space-time. The systems which has these
properties is called ''nonextensive'' and if a system does obey these
restrictions, BG statistics seems to be inappropriate and one might need a
nonextensive formalism to study on the physical system.

It will be useful to write the pseudo-additivity entropy rule, 
\begin{equation}
S_{q}(A+B)/k=\left( S_{q}(A)/k\right) +\left( S_{q}(B)/k\right) +(1-q)\left(
S_{q}(A)/k\right) \left( S_{q}(B)/k\right) ,
\end{equation}
which reflects the character of the entropic index $q$ and also of the
nonextensivity. In this equation, $A$ and $B$ are two independent systems.
Since all cases $S_{q}\geqslant 0$ (which is called nonnegativity property), 
$q<1$, $q=1$ and $q>1$, correspond to superadditivity , additivity and
subadditivity respectively.

Because of some unfamiliar consequences of first two energy constraints
discussed in [13,19], Tsallis et al. present the third internal energy
constraint as follows: 
\begin{equation}
\frac{\sum_{i=1}^{W}p_{i}^{q}\varepsilon _{i}}{\sum_{i=1}^{W}p_{i}^{q}}%
=U_{q}{}^{(3)}.
\end{equation}
This choice is commonly considered to study physical systems and is denoted
as the Tsallis-Mendes-Plastino (TMP) choice. The optimization of Tsallis
entropy given by Eq.(35) according to the third choice of the energy
constraint results in 
\begin{equation}
p_{i}^{(3)}=\frac{\left[ 1-(1-q)\beta (\varepsilon
_{i}-U_{q}^{(3)})/\sum_{j=1}^{W}(p_{j}^{(3)})^{q}\right] ^{\frac{1}{1-q}}}{%
Z_{q}^{(3)}}
\end{equation}
with 
\begin{equation}
Z_{q}^{(3)}=\sum_{i=1}^{W}\left[ 1-(1-q)\beta (\varepsilon
_{i}-U_{q}^{(3)})/\sum_{j=1}^{W}(p_{j}^{(3)})^{q}\right] ^{\frac{1}{1-q}}.
\end{equation}
Therefore the $q$-expectation value of any observable is defined as 
\begin{equation}
\left\langle A_{i}\right\rangle _{q}=\frac{\sum_{i=1}^{W}p_{i}^{q}A_{i}}{%
\sum_{i=1}^{W}p_{i}^{q}},
\end{equation}
where $A$ represents any observable quantity which commutes with
Hamiltonian. As can be expected, when $q=1$, the $q$-expectation value of
the observable reduces to the conventional one. At this stage, the important
point is to solve Eq.(40) which is an implicit one, and Tsallis et al.
suggest two different approaches; ''$iterative\,\,procedure$'' and ''$\beta
\rightarrow \beta 
{\acute{}}%
$'' transformation. The nonextensivity, in fact, appears in systems where
long range interactions and/or fractality exist, and such properties have
been invoked in recent models of manganites [20,21] as well as in the
interpretation of experimental results. These properties are also appear in
[22] where the role of the competition between different phases to the
physical properties of these materials is emphasized. The formation of
micro-clusters of competing phases, with fractal shapes, randomly
distributed in the material is considered in [23,24], and the role of long
range interactions to phase segregation [25,26].

If we look at the recent studies in which TT is employed, we see also that
the entropic index is frequently established from the dynamics of the system
under consideration analytically. Some of such studies are: Boghosian et
al.'s study [27] for the incompressible Navier-Stokes equation in Boltzmann
lattice models, Baldovin et al.'s studies [28,29] at the edge of chaos for
the logistic map universality class, Oliveira et al.' studies [20,21] on
manganites.

According to these axioms of TT given above, the order parameters are
calculated by using Eq.(40) instead of Eqs.(30) and (31). The exponential
form of any function, say x, is given by 
\begin{equation}
\exp _{q}(x)=\left[ 1+(1-q)x\right] ^{\frac{1}{1-q}}
\end{equation}
in Tsallis thermostatistics. Substituting this equation in the calculation,
Eq.(40) is employed and then the self-consistent equations are solved to
obtain the order parameters.

The generalized free energy then follows within TT: 
\begin{equation}
(F_{m})_{q}=\frac{RT}{2}\left[ a\,\left( \overline{d_{0,0}^{(2)}}\right)
_{q}+b\,\left( \overline{d_{0,2}^{(2)}\cos (2\gamma )}\right) _{q}\right]
-RT\ln _{q}(Z_{q}),
\end{equation}
where $ln_{q}(x)=(Z_{q}^{(1-q)}-1)/(1-q)$.

Now it is the time to give the results of application of TT to the Luckhurst
et al.'s theory presented above.

\section{Results and Discussion}

The first stage in our calculation is the identification of the transition
from the uniaxial nematic phase to the isotropic phase. This may be
accomplished for a given molecular symmetry by determining the value of $%
\alpha $\ at which the orientational free energy vanishes, provided there is
no volume change at the transition. The results of such calculations are
listed Tables (1-3). Table 1 shows the transition temperatures, the critical
values of order parameters $\overline{d_{0,0}^{(2)}}$ and $\overline{%
d_{0,2}^{(2)}\cos (2\gamma )}$ at the transition temperature and the entropy
changes at the transition concerning $\lambda =0.1$ for some values of $q$.
We can see immediately that increasing the entropic index increases $\alpha
_{K}$ and so decreases the nematic-isotropic transition temperature for
constant $c_{200}$. Similar behaviour is also seen from Tables (2) and (3)
concerning $\lambda =0.2$ and $0.3$ respectively. Also we see from Tables
(1-3) that increasing the entropic index increases the order parameter $%
\overline{d_{0,0}^{(2)}}$ at the transition for $\lambda =0.1,0.2$ and $0.3$%
. However there is a somewhat suprising result that as $q$ is increased, the
secondary order parameter $\overline{d_{0,2}^{(2)}\cos (2\gamma )}$ also
increases at the transition for $\lambda =0.2,0.3$ while this is not the
case for $\lambda =0.1$, that is, the secondary order parameter decreases
with increasing $q$ for $\lambda =1$. We also observe from the Tables (1-3)
that the entropy changes at the transition increases with increasing $q$.

In Fig.(1), we illustrate the dependence of the order parameter $\overline{%
d_{0,0}^{(2)}}$ and on reduced temperature for some $q$ values, with $%
\lambda =0.1$ only. The results are plotted as a function of the reduced
variable $\left( \alpha /\alpha _{K}\right) $\ \ which is identical to the
reduced temperature $T/T_{K}$ provided the coefficients $c_{2jp}$ are
themselves independent of temperature.\ We observe that as $q$ increased,
the transition becomes more markedly first-order. It is well known from [10]
that the effect of introducing deviations from cylindrical symmetry is to
lower the curves for $\lambda =0,0.1,0.2$ and $0.3$ respectively, as well as
changing their slopes. It is clear from Fig.(1) that as $q$ is decreased,
the similar behaviour is also seen, that is, the slope of the curve for $%
\lambda =0.1$ decreases with increasing the entropic index $q$. The similar
behaviour could also expected for other $\lambda $\ values $(\lambda
=0,0.1,0.3)$. The secondary order parameter $\overline{d_{0,2}^{(2)}\cos
(2\gamma )}$ is plotted as a function of $\overline{d_{0,0}^{(2)}}$, that is 
$\overline{P_{2}}$, in Fig.(2) with $\lambda =0.2$ for various values of the
entropic index. These results exhibit an unusual behaviour, for the order
parameter $\overline{d_{0,2}^{(2)}\cos (2\gamma )}$\ is observed to increase
with increasing $\overline{P_{2}}$\ pass through a maximum and then
decrease. It is interesting to note that as $q$ is decreased, the maximum
value of the secondary order parameter assumes a higher value than those
concerning lower value of $q$.

Another important point is that Maier-Saupe theory gives a universal value
of the $\overline{d_{0,0}^{(2)}}$ at the transition temperature $((\overline{%
P_{2}})_{c}=0.429)$. However both Luckhurst et al.'s theory and the
generalized form of the Maier-Saupe theory, including the present study,
assume different values of this order parameter with changing $\lambda $ and 
$q$ parameters. This fact is consistent with experiment in that experimental
values of $\overline{d_{0,0}^{(2)}}$ for various nematics assume in a range
of $0.25-0.5$ [30]. While $\lambda $ denotes the ratio $c_{220}/c_{200}$,
i.e. it is responsible from the deviations from the cylindrical symmetry, $q$
is a measure of the nonextensivity of the system. But it is interesting that
they exhibit some similar behaviour about the critical value of the order
parameter $\overline{d_{0,0}^{(2)}}$ at the transition.

\begin{tabular}{|c|c|c|c|c|}
\hline
$
\begin{array}{c}
\lambda =0.1 \\ 
q
\end{array}
$ & $\alpha _{K}$ & $\overline{P_{2}}$ & $\overline{D_{0,2}^{2}}$ & $%
S_{c}/Nk $ \\ \hline
$0.99$ & $4.458$ & $0.396$ & $0.01719$ & $0.348$ \\ \hline
$0.995$ & $4.481$ & $0.402$ & $0.01716$ & $0.357$ \\ \hline
$1$ & $4.505$ & $0.408$ & $0.0171$ & $0.384$ \\ \hline
$1.005$ & $4.528$ & $0.413$ & $0.017$ & $0.398$ \\ \hline
$1.01$ & $4.551$ & $0.417$ & $0.0169$ & $0.408$ \\ \hline
\end{tabular}

Table 1.

$
\begin{tabular}{|c|c|c|c|c|}
\hline
$
\begin{array}{c}
\lambda =0.2 \\ 
q
\end{array}
$ & $\alpha _{K}$ & $\overline{P_{2}}$ & $\overline{D_{0,2}^{2}}$ & $%
S_{c}/Nk $ \\ \hline
$0.99$ & $4.335$ & $0.332$ & $0.0334$ & $0.256$ \\ \hline
$0.995$ & $4.36$ & $0.336$ & $0.0335$ & $0.265$ \\ \hline
$1$ & $4.386$ & $0.341$ & $0.0336$ & $0.275$ \\ \hline
$1.005$ & $4.412$ & $0.348$ & $0.0337$ & $0.301$ \\ \hline
$1.01$ & $4.438$ & $0.354$ & $0.0338$ & $0.314$ \\ \hline
\end{tabular}
$

Table 2.

$
\begin{tabular}{|c|c|c|c|c|}
\hline
$
\begin{array}{c}
\lambda =0.3 \\ 
q
\end{array}
$ & $\alpha _{K}$ & $\overline{P_{2}}$ & $\overline{D_{0,2}^{2}}$ & $%
S_{c}/Nk $ \\ \hline
$0.99$ & $4.104$ & $0.2$ & $0.0399$ & $0.101$ \\ \hline
$0.995$ & $4.132$ & $0.203$ & $0.0404$ & $0.105$ \\ \hline
$1$ & $4.160$ & $0.207$ & $0.041$ & $0.112$ \\ \hline
$1.005$ & $4.1894$ & $0.212$ & $0.0417$ & $0.117$ \\ \hline
$1.01$ & $4.218$ & $0.218$ & $0.0424$ & $0.126$ \\ \hline
\end{tabular}
$

Table 3.

\section{Conclusion}

Many molecular theories assume that an essential characteristic of compounds
forming liquid crystals is the rod-like shape of their constituent molecules
that means an high length to breadth ratio. This assumption leads to the
approximation that the molecules might be assumed to have cylindrical
symmetry. However the molecules, in fact, are lath-like and thus do not
possess the high symmetry. In this sense, a molecular theory was developed
[10] for an ensemble of such particles based upon a general expansion of the
pairwise intermolecular potential together with the molecular field
approximation.

Tsallis thermostatistics has been commonly employed in studying on the
physical systems as\ a nonextensive statistics and many studies are
investigating the nonextensive effects in different systems and phenomena.
Thus we would like to handle, in this study, this molecular field theory
[10] and to investigate the effects of the nonextensivity for uniaxial
nematics formed by non-cylindrically symmetric molecules. With this aim, we
investigate the dependence of the long range order parameter on reduced
temperature and report the variation of the critical values of the order
parameters and entropy change at the transition temperature with the
entropic index.

\newpage

\section*{References}

[1] A. Saupe, Z. Naturf. A. 19 (1964) 161; P. Diehl and C. L. Khetrapal, NMR
Basic Principles Prog. 1 (1969) 1.

[2] J. A. Pople, Proc. R. Soc. A 221 (1954) 498.

[3] R. Alben, J. R. McColl, C. S. Shih, Solid St. Commun. 11 (1972) 1081.

[4] Niederberger, W., Diehl, P., Lunazzi, L., Molec. Phys. 26 (1973) 571.

[5] M. J. Freiser, Phys. Rev. Lett. 24 (1970) 1041; Liquid Crystals 3, 1972,
edited by G. H. Brown and M. M. Labes, Part 1, 281.

[6] C. S. Shih, R. Alben, J. Chem. Phys. 57 (1972) 3057.

[7] R. Alben, Phys. Rev. Lett. 30 (1973) 778.

[8] J. P. Straley, Phys. Rev. A 10 (1974) 1881.

[9] E. Sackman, P. Krebs, J. U. Rega, J. Voss, H. M\"{o}hwald, Molec.
Crystals and Liq. Crystals 24 (1973) 283.

[10] G. R. Luckhurst, C. Zannoni, P. L. Nordio, U. Segre, Mol. Phys. 30
(1975) 1345.

[11] O. Kayacan, F. B\"{u}y\"{u}kk\i l\i \c{c}, D. Demirhan, Physica A 301
(2001) 255.

[12] O. Kayacan, Physica A 328 (2003) 205.

[13] O. Kayacan, Chem. Phys. in press.

[14] M. E. Rose, Elementary Theory of Angular Momentum, John Wiley \& Sons,
1957.

[15] T. D. Shultz, Liquid Crystals, 3, 1972, edited by G. H. Brown and M. M.
Labes, Part 1, 263.

[16] A. Saupe, Angew. Chem. Int. Edition 7 (1968) 97.

[17] see http://tsallis.cat.cbpf.br for an updated bibliography.

[18] C. Tsallis, J. Stat. Phys. 52 (1988) 479.

[19] C. Tsallis, R.S. Mendes and A.R. Plastino, Physica A (1998) 534.

[20] M. S. Reis, J. P. Ara\~{u}jo, V. S. Amaral, E. K. Lenzi, and I. S.
Oliveira, Phys. Rev. B 66 (2002) 134417.

[21] M.S. Reis, J.C.C. Freitas, M.T.D. Orlando, E.K. Lenzi, I.S. Oliveira,
Europhys. Lett. 58 (2002) 42.

[22] E. Dagotto, T. Hotta, A. Moreo, Phys. Rep. 344 (2001) 1.

[23] M. Mayr, A. Moreo, J.A. Verges, J. Arispe, A. Feiguin, E. Dagotto,
Phys. Rev. Lett. 86 (2001) 135.

[24] A.L. Malvezzi, S. Yunoki, E. Dagotto, Phys. Rev. B 59 (1999) 7033.

[25] A. Moreo, S. Yunoki, E. Dagotto, Science 283 (1999) 2034.

[26] J. Lorenzana, C. Castellani, C.D. Castro, Phys. Rev. B 64 (2001) 235127.

[27] B.M. Boghosian, P.J. Love, P.V. Coveney, I.V. Karlin, S. Succi, J.
Yepez, ''Galilean-invariant lattice-Boltzmann models with H-theorem'',
preprint (2002), [e-print: cond-mat/0211093].

[28] F. Baldovin, A. Robledo, Phys. Rev. E 66 (2002) 045104 (R).

[29] F. Baldovin, A. Robledo, ''Nonextensive Pesin identity: exact
renormalization group analytical results for the dynamics at the edge of
chaos of the logistic map'', preprint (2003), [e-print: cond-mat/0304410].

[30] A. Beguin, J. C. Dubois, P. Le Barny, J. Billard, F. Bonamy, J. M.
Busisine, P. Cuvelier, Sources of thermodynamic data on mesogens, Mol.
Cryst. Liq. Cryst. 115 (1984) 1.

\newpage

\section{Figure and Table Captions \newline
}

Table 1. The order parameters and entropy change at the nematic-isotropic
transition with $\lambda =0.1$ for some $q$ values.\bigskip

Table 2. The order parameters and entropy change at the nematic-isotropic
transition with $\lambda =0.2$ for some $q$ values.\bigskip

Table 3. The order parameters and entropy change at the nematic-isotropic
transition with $\lambda =0.3$ for some $q$ values.

\bigskip

Figure 1. The variation of the order parameters $\overline{d_{0,0}^{(2)}}$
and $\overline{d_{0,2}^{(2)}\cos (2\gamma )}$ with the reduced temperature
for some $q$ values, with $\lambda =0.2$. $q=1$ represents the Luckhurst et
al.'s theory.

Figure 2. The dependence of $\overline{d_{0,2}^{(2)}\cos (2\gamma )}$ on $%
\overline{d_{0,0}^{(2)}}(=\overline{P_{2}})$ for various $q$ values.

\end{document}